\documentclass[12pt, a4paper]{article}
\usepackage[utf8]{inputenc}
\usepackage{dcolumn,lscape}
\usepackage{amsmath,longtable,multicol,dcolumn,tabularx,graphicx,amssymb}
\usepackage{exscale,amsthm,multirow,rotating,endfloat}
\usepackage{natbib}
\usepackage{bbm}
\usepackage{tikz,dsfont,float}
\usepackage[caption = false]{subfig}
\newcommand{\xvec}{\boldsymbol}
\newcommand{\xmat}{\mathbf}
\newcommand{\xset}{\mathds}

\newtheorem{theorem}{Theorem}
\newtheorem{example}{Example}

\newtheorem{corollary}{Corollary}
\newtheorem{lemma}{Lemma}

\begin{document}

\def\spacingset#1{\renewcommand{\baselinestretch}%
	{#1}\small\normalsize} \spacingset{1}


\title{\bf Generalized Spatial and Spatiotemporal ARCH Models}
\author{Philipp Otto\\
	\small{Leibniz University Hannover, Germany}\\
	Wolfgang Schmid\\
	\small{European University Viadrina, Frankfurt (Oder), Germany}}
\maketitle

\begin{abstract}
	In time-series analyses, particularly for finance, generalized autoregressive conditional heteroscedasticity (GARCH) models are widely applied statistical tools for modelling volatility clusters (i.e., periods of increased or decreased risk). In contrast, it has not been considered to be of critical importance until now to model spatial dependence in the conditional second moments. Only a few models have been proposed for modelling local clusters of increased risks. In this paper, we introduce a novel spatial GARCH process in a unified spatial and spatiotemporal GARCH framework, which also covers all previously proposed spatial ARCH models, exponential spatial GARCH, and time-series GARCH models. In contrast to previous spatiotemporal and time series models, this spatial GARCH allows for instantaneous spill-overs across all spatial units. For this common modelling framework, estimators are derived based on a non-linear least-squares approach. Eventually, the use of the model is demonstrated by a Monte Carlo simulation study and by an empirical example that focuses on real estate prices from 1995 to 2014 across the ZIP-Code areas of Berlin. A spatial autoregressive model is applied to the data to illustrate how locally varying model uncertainties (e.g., due to latent regressors) can be captured by the spatial GARCH-type models.
\end{abstract}

\noindent%

\spacingset{1.45} 

	\section{Introduction}\label{sec:introduction}

	Recent literature have dealt with the extension of generalized autoregressive conditional heteroscedasticity (GARCH) models to spatial and spatiotemporal processes (e.g. \citealt{Otto16_arxiv,Otto18_spARCH,Otto19_statpapers,Sato17,Sato18b,Sato18a}). Whereas the classical ARCH model is defined as a process over time, these random processes have a multidimensional support. Thus, they allow for spatially dependent second-order moments, while the local means are uncorrelated and constant in space (see \citealt{Otto19_statpapers}). \cite{Sato17,sato2020spatial} introduced a random process incorporating elements of GARCH and exponential GARCH (E-GARCH) processes, which is, however, neither a GARCH nor an E-GARCH process. Moreover, \cite{Otto16_arxiv} only focussed on spatial ARCH processes without considering the influences from the realized, conditional variance at neighbouring locations. Direct extensions of GARCH and E-GARCH processes to spatial settings do not exist among current research.
	
	{In this paper, we introduce a completely novel generalized spatial ARCH model (spGARCH). Because a general definition of this model is used, time-series GARCH models (\citealt{Bollerslev86}), the previously introduced spatial ARCH (\citealt{Otto16_arxiv}) and the hybrid GARCH (\citealt{Sato17,Sato18a}) are included. This definition also allows us to define a spatial logarithmic and exponential spatial GARCH model, which will be the subject of a future paper. Moreover,} other GARCH-type models, like threshold or multivariate GARCH models, can easily be constructed. 
	This unified spatial GARCH process is a completely new class of models in spatial econometrics, {for which we derive consistent estimators based on a non-linear least-squares approach.} In addition, all models are computationally implemented in one library, the R-package \texttt{spGARCH} {(version $> 2.0$)}.
	
	From a practical perspective, this unified spatial GARCH model can be used to model spill-over effects in the conditional variances {across the spatial units}. That means that an increasing variance in a certain region of the considered space would lead to an increase or decrease in the adjacent regions, depending on the direction (sign) of the spatial dependence. {Compared to previous spatiotemporal GARCH models, these spill-overs are instantaneous.} Local climate risks, such as fluctuations in the temperature and precipitation, or financial risks in spatially constrained markets, such as real estate or labour, could be modelled using this approach. Furthermore, spatial GARCH-type models can be used as error models for any linear or non-linear spatial regression model to account for local model uncertainties (i.e., areas in which the considered models perform worse than in others). Such model uncertainties can be considered to be a kind of local risk.

	The remainder of this paper is structured as follows. In the next section, we introduce the {generalized framework of spatial and spatiotemporal autoregressive conditional heteroscedasticity models} and discuss two examples nested within this approach, more precisely, the novel spatial GARCH (as an equivalent to the time-series GARCH models) and the hybrid spatial GARCH processes by \cite{Sato17,sato2020spatial}. Following from there, a non-linear least-squares procedure is introduced for this model class. These theoretical sections are followed up with a discussion of the insights gained from simulation studies. The paper then supplies a real-world example, namely the real estate prices in the German capital city of Berlin. In Section \ref{sec:conclusion}, we stress some important extensions for future research before concluding the paper.

	\section{Spatial and Spatiotemporal GARCH-Type Models}\label{sec:models}

	Let $\left\{Y(\xvec{s}) \in \xset{R}: \xvec{s} \in D_{\xvec{s}} \right\}$ be a univariate stochastic process, where $D_{\xvec{s}}$ represents a set of possible locations in a $q$-dimensional space. Thus, spatial and spatiotemporal models are both covered by this approach. With regards to spatiotemporal processes, the temporal dimension can be easily considered as one of the $q$ dimensions. In addition, time-series GARCH models are included for $q = 1$.
	
	Let $\xvec{s}_1, \ldots, \xvec{s}_n$ denote all locations, and let $\xvec{Y}$ stand for the vector of observations $\left(Y\left(\xvec{s}_i\right)\right)_{i = 1, \ldots, n}$. The commonly applied spatial autoregressive (SAR) model implies that the conditional variance $Var(Y(\xvec{s}_i) | Y(\xvec{s}_j), j \neq i)$ is constant (cf. \citealt{Cressie93,Cressie11}) and does not depend on the observations of neighbouring locations. This approach is extended by assuming the changes in the volatility can spill over to neighbouring regions and that conditional variances can vary over space, resulting in clusters of high and low variance.
	As in time-series ARCH models developed by \cite{Engle82}, the vector of observations is given by the non-linear relationship
	\begin{equation}
		\xvec{Y} = \text{diag}(\xvec{h})^{1/2} \xvec{\varepsilon} \, \label{eq:initial}
	\end{equation}
	where $\xvec{h} = (h(\xvec{s}_1), \ldots, h(\xvec{s}_n))'$ and $\xvec{\varepsilon} = (\varepsilon(\xvec{s}_1), \ldots, \varepsilon(\xvec{s}_n))'$ is a noise component, which is later specified in more detail.
	
	{Moreover}, we assume that a known function $f$ exists, which relates $\xvec{h}$ to a vector $\xvec{F} = (f(h(\xvec{s}_1)), \ldots, f(h(\xvec{s}_n)))^\prime$. {This general approach is beneficial because} different spatial GARCH-type processes can be defined by choosing $f$ and a suitable model of $\xvec{F}$. For instance, they could have additive or multiplicative dynamics, or the spill-over effects in the conditional variances could be global or locally constrained to direct neighbouring observations. {In this paper, we initially focus on generalized spatial GARCH models (spGARCH), which are analogously defined to the time-series GARCH models and have additive dynamics (cf. \citealt{Bollerslev86}). Besides, previously introduced spatial ARCH models are nested within this general approach} (e.g., \citealt{Otto18_spARCH,sato2020spatial}; see also Examples \ref{example:spARCH} and \ref{example:spHARCH} in Section \ref{sec:examples}).

	\subsection{{Generalized Approach}}
	
	{Below,} we introduce a general approach covering some important spatial and spatiotemporal GARCH-type models, namely the spatial ARCH model of \cite{Otto16_arxiv,Otto18_spARCH} and the hybrid model of \cite{sato2020spatial}. For these models, vector $\xvec{F}$ is chosen as
	\begin{equation}\label{eq:unified}
		\xvec{F} = \xvec{\alpha} + \xmat{W}_1 \xvec{\gamma}(\xvec{Y}^{(2)}) + \xmat{W}_2 \xvec{F}
	\end{equation}
	with a measurable function $\xvec{\gamma}(\xvec{x}) = (\gamma_1(\xvec{x}), \ldots, \gamma_n(\xvec{x}))^\prime$ and \linebreak
	$\xvec{Y}^{(2)} = (Y(\xvec{s}_1)^2, \ldots , Y(\xvec{s}_n)^2)^\prime$. The weighting matrices $\xmat{W}_1 = (w_{1,ij})_{i,j = 1, \ldots, n}$ and $\xmat{W}_2 = (w_{2,ij})_{i,j = 1, \ldots, n}$ are assumed to be non-negative with zeros on the diagonal (i.e., $w_{v,ij} \ge 0$ and $w_{v,ii}=0$ for all $i,j = 1, \ldots, n$ and $v = 1, 2$). Moreover, let $\xvec{\alpha} = (\alpha_i)_{i = 1, \ldots, n}$ be a positive vector.
	
	First, we discuss under what conditions the process is well-defined. {To do this, we make use of the Banach fixed point theorem for random processes. The field of random fixed point theorems has been {studied} by several authors (e.g., \citealt{hanvs1957reduzierende,bharucha1976fixed,tan1997random}).
		
	In the following, we make use of the notation $\xmat{E} = \mbox{diag}(\varepsilon(\xvec{s}_1)^2, \ldots, \varepsilon(\xvec{s}_n)^2)$. {Considering} the operator
		\begin{equation}\label{eq:T}
			\xvec{T}(\omega)\circ \xvec{z} = \xvec{\alpha} + \xmat{W}_1 \xvec{\gamma}(\xmat{E}(\omega) \xvec{z}) - (\xmat{I} - \xmat{W}_2) ( f(z_i) )_{i=1,...,n} + \xvec{z}
		\end{equation}
		defined on $I\!\!R^n$ with a norm $||.||${, the following conditions can be derived such that the process is well-defined.}
		
		\begin{theorem}\label{solution}
			Suppose that the operator $\xvec{T}(\omega)$ defined in (\ref{eq:T}) is a continuous random operator on $(I\!\!R^n, ||.||)$ to itself and that there is a a non-negative real-valued random variable $L_n(\omega) < 1$ a.s. such that  $|| \xvec{T}(\omega) \circ \xvec{z}_1 - \xvec{T}(\omega) \circ \xvec{z}_2|| \le L_n(\omega)  ||\xvec{z}_1 - \xvec{z}_2||$ for all $\xvec{z}_1, \xvec{z}_2 \in I\!\!R^n$. Then the equations (1) and (2) have exactly one real-valued measurable solution $\xvec{z}$.
		\end{theorem}
		
		{The proof of this theorem is given in the Appendix.} Note that the condition (\ref{eq:T}) is fulfilled if, for example, $\xvec{\gamma}$ satisfies a Lipschitz condition with constant $L_1$, $(f(z_i))_{i=1,...,n}$ satisfies a Lipschitz condition with a constant $L_2$ and $L_n := L_1 ||\xmat{W}_1 \xmat{E}|| + L_2 ||\xmat{I} - \xmat{W}_2|| < 1$ where we make use of the matrix norm induced by the vector norm. However, in order to guarantee that $L_n$ does not depend on $n$, we need stronger conditions. If we take the 1-norm and if the matrices $\xmat{W}_1$ and $\xmat{W}_2$ are row-standardized then $||\xmat{I} - \xmat{W}_2|| < 2$. To ensure that $||\xmat{W}_1 \xmat{E}||$ is bounded we have to assume that the $\varepsilon(\xvec{s}_i)$ are uniformly bounded.  We refer to \cite{Otto18_spARCH} where this problem is discussed for a spatial ARCH process in more detail.
		
		{Moreover, it is important to note that} the operator $\xvec{T}(\omega)$ is continuous if $f$ and $\gamma$ are continuous.
		
		Further, the solution of (\ref{eq:T}) which reflects $\xvec{h}$ should be non-negative such that the process $\xvec{Y}$ is a well-defined real-valued process. In many applications, the functions $f$ and $\xvec{\gamma}$ are defined to be zero for negative values{, such that $\xvec{h}$ is always positive if $\xvec{\alpha} > 0$.} We will come back to this point later.
	}
	In addition, the fixed-point theorem of Banach implies that the sequence $\xvec{z}_m = \xvec{T}(\xvec{z}_{m-1})$, $m \ge 1$ converges to $\xvec{h}$ for given $\xvec{\alpha}$, $\xmat{E}$, $\xmat{W}_1$, and $\xmat{W}_2$. Consequently, this result {represents one way} to simulate such a process.
	
	\subsection{{Properties of spatial GARCH models}}
	
	Below, we discuss some important properties of this process including the following condition for stationarity.

	\begin{corollary}\label{cor:stationarity2}
		Suppose that the assumptions of Theorem \ref{solution} are fulfilled {and that the solution of (\ref{eq:T}) is non-negative}. If $(\varepsilon(\xvec{s}_1),\ldots,\varepsilon(\xvec{s}_n))^\prime$ is strictly stationary, then $(Y(\xvec{s}_1), \ldots, Y(\xvec{s}_n))^\prime$ is strictly stationary as well.
	\end{corollary}
	
	Moreover, the observations $Y(\xvec{s})$ are uncorrelated with a mean of zero, as we will show in the following theorem. Thus, spatial GARCH models are suitable error models for use with other linear or non-linear spatial regression models, such as spatial autoregressive or spatial error models (see also \citealt{Elhorst10}), without affecting the mean equation. In this way, locally varying model uncertainties can be captured.

	\begin{theorem}\label{th:moments1}
		Let $i \in \{1,\ldots,n\}$. Suppose that the assumptions of Theorem \ref{solution} are satisfied {and that the solution of (\ref{eq:T}) is non-negative}. Further let $\xvec{\varepsilon}$ be sign-symmetric, i.e.,
		\[ \xvec{\varepsilon} \stackrel{d}{=} ((-1)^{v_1} \varepsilon(\xvec{s}_1),\ldots,(-1)^{v_n} \varepsilon(\xvec{s}_n)) \qquad \mbox{for all} \qquad v_1,\ldots,v_n \in \{0,1\} . \]
		\begin{itemize}
			\item[a)] Then $Y(\xvec{s}_i)$ is a symmetric random variable. All odd moments and all conditional odd moments of $Y(\xvec{s}_i)$ are zero, provided that they exist.
			\item[b)] It holds that $\mbox{Cov}(Y(\xvec{s}_i), Y(\xvec{s}_j)) = 0$ for $i \neq j$ if the second moment exists.
		\end{itemize}
	\end{theorem}
	
	In the spatial setting, however, the conditional variance $Var(Y(\xvec{s}_i) | Y(\xvec{s}_j), j \neq i)$ is not exactly equal to $h(\xvec{s}_i)$  (see \citealt{Otto19_statpapers}). Nevertheless, the interpretation of $\xvec{h}$ is similar to the conditional variance. In locations $\xvec{s}$, where $h(\xvec{s})$ is large, the conditional variance is also large and vice versa (see \citealt{Otto19_statpapers}, Fig. 1). That means that the local risk or level of uncertainty of this particular region is high compared to its neighbours. Such regions could be identified via $\xvec{h}$; this could be of interest in terms of the valuation of real estate or other immovable assets since it provides insights into an individual location's risk. 
	
	In addition to this, the spatial GARCH coefficients measure potential risk spill-overs from neighbouring locations. It is worth noting that in the case of directional spatial processes, $\xvec{h}$ is equal to the conditional variances. Thus, it can be interpreted in the same way as with time-series GARCH models (see \citealt{Otto19_statpapers}).
	
	\subsection{{Examples of spatial GARCH models}}\label{sec:examples}
	
	This {general framework} allows for a large range of GARCH-type models. Depending on the definition of $f$ and $\xvec{\gamma}$, the resulting spatial GARCH-type models have different stochastic properties. We discuss some important special cases below, starting with the spatial ARCH model (\citealt{Otto16_arxiv,Otto18_spARCH,Otto19_statpapers}), which is a direct extension of the ARCH process of \cite{Engle82} to spatial and spatiotemporal processes. It was originally introduced by \cite{Otto16_arxiv}. For more details on its stochastic properties, we refer to \citealt{Otto19_statpapers}.

	\begin{example}[Spatial ARCH process of \citealt{Otto16_arxiv}]\label{example:spARCH}
		Choosing $f(x) = x I_{[0,\infty)}(x)$, $\gamma_i(\xvec{x}) = x_i I_{[0,\infty)}(x_i)$ for $i = 1, \ldots, n$, and $\xmat{W}_2 = \xmat{0}$ the spatial ARCH (spARCH) process is obtained. It is given by
		\begin{equation*}
			Y(\xvec{s}_i) = \sqrt{h(\xvec{s}_i)} \varepsilon(\xvec{s}_i), \quad i=1,...,n
		\end{equation*}
		with
		\begin{equation*}
			\xvec{h} = \xvec{\alpha} + \xmat{W}_1 \xvec{Y}^{(2)}\, .
		\end{equation*}
	\end{example}
	
	{The process is well-defined if $||\xmat{W}_1 \xmat{E}|| < 1$. This is an immediate consequence of Theorem \ref{solution}.} Indeed, the spatial ARCH process can be easily extended to a spatial GARCH process by considering the realized values of $h(\cdot)$ in adjacent locations. {This novel spatial GARCH process is defined in the following example.}

	\begin{example}[Spatial GARCH process]\label{example:spGARCH}
		Taking {$f(x) = x I_{[0,\infty)}(x)$ and  $\gamma_i(\xvec{x}) = x_i I_{[0,\infty)}(x_i)$} for $i=1,...,n$ a spatial  GARCH (spGARCH) process is obtained. That is,
		\begin{equation*}
			Y(\xvec{s}_i) = \sqrt{h(\xvec{s}_i)} \varepsilon(\xvec{s}_i), \quad i=1,...,n
		\end{equation*}
		with
		\begin{equation*}
			\xvec{h} = \xvec{\alpha} + \xmat{W}_1 \xvec{Y^{(2)}} + \xmat{W}_2 \xvec{h} \, .
		\end{equation*}
	\end{example}

	Since $\xvec{Y}^{(2)} = \xmat{E} \xvec{h}$, the quantity $\xvec{h}$ can be specified as
	\begin{equation}\label{epsGARCH}
		\xvec{h} = ( \xmat{I} - \xmat{W}_1 \xmat{E} - \xmat{W}_2)^{-1} \xvec{\alpha} \, ,
	\end{equation}
	if the inverse exists.  For this simple example, there is a unique solution if {$||\xmat{W}_1 \xmat{E} + \xmat{W}_2|| < 1$}, as it is already expressed in Theorem \ref{solution}. {Alternatively, the condition is fulfilled if the process is directional. In this case, $\xmat{W}_1$ and $\xmat{W}_2$ are lower or upper triangular matrices (cf. \citealt{Basak18,otto2019stochastic,merk2021directional}).}

	Contrary to this approach, \cite{Sato17,Sato18b,sato2020spatial} have considered a slightly different choice of $\xvec{h}$, and have used the log-transformation to avoid any non-negativity problems of $\xvec{h}$. Thus, their model combines the GARCH and the E-GARCH attempts. For that reason, we will {refer to} it as the hybrid model {(H-spGARCH)}. Let $\xvec{h}_L = ( \log(h(\xvec{s}_i)) )_{i=1,...,n}$ and $\xvec{Y}^{(2)}_L = ( \log(Y(\xvec{s}_j)^2 ) )_{i=1,...,n}$.
	
	\begin{example}[Hybrid spatial GARCH process of \citealt{Sato17}]\label{example:spHARCH}
		Choosing $f(x) = \log(x)$ and $\gamma_i(\xvec{x}) = \log(x_i)$ the hybrid spatial GARCH (H-spGARCH) process is obtained, i.e.,
		\begin{equation*}
			Y(\xvec{s}_i) = \sqrt{h(\xvec{s}_i)} \varepsilon(\xvec{s}_i), \quad i=1,...,n
		\end{equation*}
		with
		\begin{equation*}
			\xvec{h}_L = \xvec{\alpha} + \xmat{W}_1  \xvec{Y}^{(2)}_L + \xmat{W}_2 \xvec{h}_L \, .
		\end{equation*}
	\end{example}

	The process has a unique solution if {$||\xmat{W}_1 + \xmat{W}_2|| < 1$.} This is {also} an immediate consequence of Theorem \ref{solution}. It is obtained by setting $\gamma(\xvec{x}) = (\log(x_i) )_{i=1,\ldots,n}$, $f(x)=\log(x)$, and considering the right side of \eqref{eq:T} to be a function of $\log(z)$. We see that the condition on the existence of a solution is much simpler than for the spGARCH process, since it only depends on the weight matrices and not on the random matrix $\xmat{E}$. This simplification is due to the fact that we have an additive decomposition of the function $\xvec{\gamma}$, i.e., $\xvec{\gamma}(\xmat{E} \xvec{z}) = \xvec{\phi}_1(\xmat{E}) + \xvec{\phi}_2(\xvec{z})$ with certain functions $\xvec{\phi}_1$ and $\xvec{\phi}_2$. This functional equation is solved by  the logarithm function. However, the behavior of the H-spGARCH is different to that of the spGARCH. Thus, the one or the other could be preferable for empirical applications.

	\section{Statistical Inference}
	
	In the following section, we firstly discuss the choice of the weight matrices in more detail. In a general setting, $\xmat{W}_1$ and $\xmat{W}_2$ have $n(n-1)$ free parameters, while only $n$ values are observed. In spatial econometrics, these matrices are therefore usually replaced with a parametric model to control the influence of adjacent regions. Alternatively, they might instead be estimated using statistical learning approaches, e.g., lasso-type estimators under the assumption of a certain degree of sparsity. {In this section of the paper, however, we will refer to a classical parametric model. For this, we develop an estimation method based on non-linear least squares estimators and show the consistency of these estimators.}

	\subsection{Choice of Weight Matrices}
	
	There is great flexibility in the choice of the weight matrices (see \citealt{Getis09} for an overview). In practice, these are usually dependent upon additional parameters and spatial locations.
	Frequently, it is assumed that $\xmat{W}_1 = \rho \xmat{W}^{*}_1$ and $\xmat{W}_2 = \lambda \xmat{W}^{*}_2$ with the predefined, known matrices $\xmat{W}^{*}_1$ and $\xmat{W}^{*}_2$. That is, $\xmat{W}^{*}_1$ and $\xmat{W}^{*}_2$ describe the structure of the spatial dependence, with the weights as a multiple of these specific matrices. In settings such as these, it is easy to test whether a random process exhibits such a spatial dependence, by testing the parameters $\rho$ and $\lambda$. As with time-series GARCH models, $\rho$ measures the extent to which a volatility shock in one region spills over to neighbouring regions, while $\rho + \lambda$ gives an impression how fast this effect will fade out in space (see, e.g., \citealt{Campbell97}). A more general approach can be obtained by choosing $\xmat{W}_{k} = \text{diag}(\rho_1, \ldots, \rho_1, \ldots, \rho_k, \ldots, \rho_k) {\xmat{W}_{\cdot}^{*}}$ as the weights for $k \in \{1,2\}$. Here, different areas are weighted in different ways. For instance, all counties of state $i$ are weighted by $\rho_i$, while counties of another state, $j$, get a different weighting factor, $\rho_j$. Alternatively, $\xmat{W}_{k}^{*}$ could be chosen as $(K_\theta(\xvec{s}_i - \xvec{s}_j))_{i,j = 1, \ldots, n}$ for $k \in \{1,2\}$ with a known function $K$. In this case, the spatial correlation depends on the distance between two locations. For instance, inverse distance weighting schemes $K(\xvec{x}) = ||\xvec{x}||^{-k}$ with $k$ being estimated, or anisotropic weighting schemes dependent upon the bearing between two locations.
	
	\subsection{Parameter Estimation}
	
	Below, we assume that the weight matrices have the structure
	\begin{equation}\label{weight}
		\xmat{W}_1 = \rho \xmat{W}_1^*, \xmat{W}_2 = \lambda \xmat{W}_2^*, \xvec{\alpha} = \alpha \xvec{1} .
	\end{equation}
	Thus, the model has three parameters to be estimated, $\xvec{\vartheta} = (\rho, \lambda, \alpha)^\prime$. Let $\xvec{\vartheta}_0$ denote the true parameters.
	In the following we use the symbol $||\xvec{x}||_2$ for the Euclidean norm of a vector $\xvec{x}$ and $||\xmat{A}||_\infty = max_{1 \le i \le n} \sum_{j=1}^{n} |a_{ij}|$ for the matrix norm of an $n \times n$ matrix $\xmat{A}$, which is induced by a maximum norm.
	
	
	One possible method for estimating the parameters is the non-linear least-squares approach (NLSE). 
	Squaring the components of \eqref{eq:initial} and taking the logarithms, we get that for $i=1, \ldots, n$
	\begin{eqnarray*}
		\mbox{log}(Y(\xvec{s}_i)^2) & = & \mbox{log}(h(\xvec{s}_i)) + \mbox{log}(\varepsilon(\xvec{s}_i)^2)\\
		& = & E( \mbox{log}(\varepsilon(\xvec{s}_i)^2) ) + \mbox{log}(h(\xvec{s}_i)) + \eta(\xvec{s}_i)
	\end{eqnarray*}
	with $\eta(\xvec{s}_i) = \mbox{log}(\varepsilon(\xvec{s}_i)^2) - E( \mbox{log}(\varepsilon(\xvec{s}_i)^2) )$. Now $\eta(\xvec{s}_i), i=1,...,n$ is a white noise process. Moreover, it follows with $\tau(x) = f(\mbox{exp}(x))$ that
	\begin{eqnarray*}
		\xvec{F} & = & ( \tau(\mbox{log}(h(\xvec{s}_i)) )_{i=1,...,n} = (\xmat{I} -  \lambda \xmat{W}_2^*)^{-1} (
		\alpha \xvec{1} + \rho \xmat{W}_1^* \xvec{\gamma}(\xvec{Y}^{(2)} ) ) \\
		& = &  (\xmat{I} -  \lambda \xmat{W}_2^*)^{-1} (\alpha \xvec{1} + \rho \xmat{W}_1^* \tilde{\xvec{\gamma}}(\mbox{log}(\xvec{Y}^{(2)}) ) )
	\end{eqnarray*}
	where $ \tilde{\xvec{\gamma}}(\xvec{x}) = ( \gamma_i(\mbox{exp}(x_1),..., \mbox{exp}(x_n)) )_{i=1,...,n}$. Note that $(\xmat{I} - \lambda \xmat{W}_2^*)^{-1}$ exists if $||\lambda \xmat{W}_2^*|| < 1$.
	
	\noindent Now, let $( c_i(\lambda) )_{i=1,...,n} = (\xmat{I} -  \lambda \xmat{W}_2^*)^{-1} \xvec{1}$ and $( \xvec{d}_i(\lambda)^\prime )_{i=1,...,n} = (\xmat{I} -  \lambda \xmat{W}_2^*)^{-1} \xmat{W}_1^*$.  In order to denote the dependence on $\xvec{\vartheta}$ we write $h_{\xvec{\vartheta}}(\xvec{s}_i)$, $i=1,...,n$. Then,
	\[ \mbox{log}(h_{\xvec{\vartheta}}(\xvec{s}_i)) = \tau^{-1}(\alpha \, c_i(\lambda) + \rho \, \xvec{d}_i(\lambda)^\prime \,  \tilde{\xvec{\gamma}}(\mbox{log}(\xvec{Y}^{(2)} ) ) ) . \]
	Here, we assume that $c = E( \mbox{log}(\varepsilon(\xvec{s}_i)^2) )$ is a known quantity. Using $H_i = \mbox{log}(Y(\xvec{s}_i)^2) - E( \mbox{log}(\varepsilon(\xvec{s}_i)^2) )$ and $\xvec{H} = ( H_i )_{i=1,...,n}$ the estimators of the parameters $\alpha$, $\lambda$, and $\rho$ are obtained by minimizing the non-linear sum of squares
	\[
	\sum_{i=1}^n \left( H_i - \mbox{log}(h_{\xvec{\vartheta}}(\xvec{s}_i) \right)^2 = \sum_{i=1}^n \left( H_i - \tau^{-1}(\alpha \, c_i(\lambda) + \rho \, \xvec{d}_i(\lambda)^\prime \,  \tilde{\xvec{\gamma}}(\xvec{H} + c \xvec{1} ) ) \right)^2 \]
	with respect to $\xvec{\vartheta}$.
	
	Although $\tau^{-1}$ is a known function, this minimization problem is complex. Thus, we will impose further assumptions which are fulfilled for all relevant special cases. We will suppose that $\xvec{\gamma}(\xvec{x}) = ( \gamma( x_i ) )$ with a known function $\gamma$. Consequently, $\tilde{\xvec{\gamma}}(\xvec{x}) = ( \tilde{\gamma}(x_i) )_{i=1,...,n}$ with $\tilde{\gamma}(x) = \gamma( \mbox{exp}(x) )$, which leads to the easier minimization of
	\begin{equation}\label{min}
		Q_n(\xvec{\vartheta} ) = \frac{1}{n} \; \sum_{i=1}^n \left( H_i - \tau^{-1}\left(\alpha \, c_i(\lambda) + \rho \, \xvec{d}_i(\lambda)^\prime \,  \left( \tilde{\gamma}( H_v + c )\right) \right)_{v=1,...,n}
		\right)^2 .
	\end{equation}
	Note that since the $(i,i)$-th element of $\xmat{W}_1^*$ is zero it follows that \linebreak $\xvec{d}_i(\lambda)^\prime \,  \left( \tilde{\gamma}( H_v + c ) \right)_{v=1,...,n}$ is no function of $H_i$.
	Minimization problems of that type have been studied in detail in, e.g., \cite{Amemiya85,Potscher97,Newey94}. There, sufficient conditions are given for the {consistency} and asymptotic normality of the resulting estimators under various conditions. Note that, in the present case, $\{ H_i \}$ is a strictly stationary process. Moreover, in most papers on this topic, the regression function is assumed to be a deterministic function depending on certain parameters. In the present case, however, it is a function depending on the observations $\{ H_i \}$ which makes the analysis of the asymptotic behaviour of the estimators much harder. Further, it must be noted that a spatial problem is present. The positions $\xvec{s}_i$ are points in a space and we need a certain distance measure between these points to assess the dependence of the observations.

	\begin{theorem}\label{th:consistency}
		Suppose that $\xvec\vartheta_0  \in \Theta = [\rho_l, \rho_u] \times  [\lambda_l, \lambda_u] \times  [\alpha_l, \alpha_u] \subseteq [0, 1) \times [0,1) \times [0,\infty)$. Let $\{ \varepsilon(\xvec{s}_i) : i \in I\!\!N \}$ be independent and identically distributed random variables with existing moment $E( (\log(|\varepsilon(\xvec{s}_1)|))^2 )$. Let $f$ be a differentiable and invertible function with $f^\prime > 0$ on $(0,\infty)$ and let $\gamma$ be a measurable function on $[0, \infty)$ with $Var(\gamma(Y(\xvec{s}_1)^2)) < \infty$. Suppose that $f^\prime(h_{\xvec{\vartheta}}(\xvec{s}_i)) \, h_{\xvec{\vartheta}}(\xvec{s}_i) \ge L > 0$ for all $i, \xvec{\vartheta} \in \Theta, \omega$. Further assume that for $\xvec\vartheta = (\rho, \lambda, \alpha) \in \Theta$
		
		\begin{equation}\label{limit1}
			\lim_{n \to \infty} \frac{1}{n} \sum_{i=1}^n \left( E( \log(h_{\xvec{\vartheta}}(\xvec{s}_i)) - E(\log(h_{\xvec{\vartheta}_0}(\xvec{s}_i)) ) \right)^2 \quad \mbox{exists} ,
		\end{equation}

		\begin{equation}\label{limit2}
			\frac{1}{n} \sum_{i=1}^n \left( \log(h_{\xvec{\vartheta}}(\xvec{s}_i)) - E( \log(h_{\xvec{\vartheta}}(\xvec{s}_i))) \right)^2 \stackrel{p}{\rightarrow} 0
		\end{equation}

		as $n$ tends to infinity and that the limit function in (\ref{limit1}) has a unique minimum at $\xvec{\vartheta}_0$. Moreover, let $\xmat{W}_1^*$ and $\xmat{W}_2^*$ be row-standardized, i.e., $\xmat{W}_1^* \xvec{1} = \xmat{W}_2^* \xvec{1} = \xvec{1}$. Then the minimization problem (\ref{min}) has a solution $\hat{\xvec\vartheta}_n$ and it holds that $\hat{\xvec\vartheta}_n \stackrel{p}{\rightarrow} \xvec\vartheta_0$ as $n \rightarrow \infty$.
	\end{theorem}
	
	Note that the solution of (\ref{min}) does not have to be unique. For more details, we refer to Section 4 of \cite{Amemiya85}.
	
	Moreover, for a spGARCH process it holds that $f^\prime(x) x = x$ and $h_{\xvec{\vartheta}}(\xvec{s}_i) \ge \alpha_l > 0$ and thus the above condition is fulfilled. For a H-spGARCH process we have that $f^\prime(x) x = 1$ and thus it is fulfilled as well. Further, for a H-spGARCH process the condition (\ref{limit1}) can be easily seen to be fulfilled since for a strictly stationary process $\{ Y(\xvec{s}_i ) \}$ the quantity $E(\log(h_{\xvec{\vartheta}}(\xvec{s}_i)))$ does not depend on $i$ at all.

	{Moreover, to prove the consistency of the local minimum, the roots of the first derivative of the sum of non-linear squares with respect to the parameters must be zero, i.e.,}
	%
	\[ \frac{\partial Q_n(\xvec{\vartheta})}{\partial \xvec{\vartheta}} = 0 . \]

	\begin{theorem}\label{th:consistency2}
		Suppose that $\xvec{\Theta}$ is an open subset of $[0, 1) \times [0,1) \times [0,\infty)$ and let $\xvec{\vartheta}_0 \in \xvec{\Theta}$.
		Let $\{ \varepsilon(\xvec{s}_i) : i \in I\!\!N \}$ be independent and identically distributed random variables with existing moment $E( (\log(|\varepsilon(\xvec{s}_1)|))^2 )$. Let $f$ be a differentiable and invertible function with $f^\prime > 0$ on $(0, \infty)$ and let $\gamma$ be a measurable function on $[0, \infty)$ with $Var(\gamma(Y(\xvec{s}_1)^2)) < \infty$. Suppose that $f^\prime(h_{\xvec{\vartheta}}(\xvec{s}_i)) \, h_{\xvec{\vartheta}}(\xvec{s}_i) \ge L > 0$ for all $i, \xvec{\vartheta} \in \Theta, \omega$. Further assume that there is an open neighbourhood $N$ of $\xvec\vartheta_0$ such that
		
		\begin{equation}\label{limit4}
			\lim_{n \to \infty} \frac{1}{n} \sum_{i=1}^n \left( E( \log(h_{\xvec{\vartheta}}(\xvec{s}_i)) - E(\log(h_{\xvec{\vartheta}_0}(\xvec{s}_i)) ) \right)^2 \quad \mbox{exists}
		\end{equation}
		
		and
		
		\begin{equation}\label{limit5}
			\frac{1}{n} \sum_{i=1}^n \left( \log(h_{\xvec{\vartheta}}(\xvec{s}_i)) - E( \log(h_{\xvec{\vartheta}}(\xvec{s}_i))) \right)^2 \stackrel{p}{\rightarrow} 0
		\end{equation}
		
		as $n$ tends to infinity and that the limit function in (\ref{limit4}) has a unique minimum in $\xvec{\vartheta}_0$. Moreover, let $\xmat{W}_1^*$ and $\xmat{W}_2^*$ be row-standardized, i.e., $\xmat{W}_1^* \xvec{1} = \xmat{W}_2^* \xvec{1} = \xvec{1}$.
		
		Let $\xvec{\Theta}_T$ denote the set of roots of the equation
		
		\[ \frac{\partial Q_n(\xvec{\vartheta})}{\partial \xvec{\vartheta}} = 0 . \]
		
		Then it holds for all $\varepsilon > 0$ that
		
		\[ \lim_{n \to \infty} P( inf_{\xvec{\vartheta} \in \Theta_T} (\xvec{\vartheta} - \xvec{\vartheta}_0)^\prime (\xvec{\vartheta} - \xvec{\vartheta}_0) > \varepsilon) = 0 . \]
		
	\end{theorem}

	
	{Like in the previous section, we will now consider a special case of the general framework, namely an spGARCH model. That is,} we choose $f(x) = x I_{ [0, \infty ]}(x)$ while $\gamma$ is an arbitrary function satisfying certain conditions.

	\begin{lemma}\label{lemma:spGARCH}
		Let $\{ Y_t \}$ be an spGARCH process. Suppose that $\xvec{\Theta} = (0,1) \times (0,1) \times (0,\infty)$.
		Let $\{ \varepsilon(\xvec{s}_i) : i \in I\!\!N \}$ be independent and identically distributed random variables with existing moment $E( (\log(|\varepsilon(\xvec{s}_1)|))^2 )$. Let $\gamma$ be a non-negative measurable function on $[0, \infty)$ with  $Var(\gamma(Y(\xvec{s}_1)^2)) < \infty$. Suppose that $\{ Y(s_i) \}$ is strictly stationary. Moreover, let $\xmat{W}_1$ and $\xmat{W}_2$ be row-standardized.
		\begin{itemize}
			\item[a)] If there is an open neighbourhood $N$ of $\xvec{\vartheta}_0$ such that for all $\xvec{\vartheta} = (\rho, \lambda, \alpha) \in N$ it holds with $\xvec{\Delta} = \left( \gamma( Y(\xvec{s}_j)^2 ) - E(  \gamma( Y(\xvec{s}_j)^2 )  \right)_{j=1,...,n}$ that
			\begin{equation}\label{limit7}  
				\frac{1}{n} \;  \xvec{\Delta}^\prime  \xmat{W}_1^{* \prime}   (\xmat{I} - \lambda \xmat{W}_2^{* \prime})^{-1}   (\xmat{I} - \lambda \xmat{W}_2^*)^{-1} \xmat{W}_1^* \xvec{\Delta} \stackrel{p}{\rightarrow} 0 
			\end{equation}
			as $n$ tends to infinity then it holds that the assumption (\ref{limit5}) is fulfilled.
			\item[b)] If
			\begin{equation}\label{limit8}
				\frac{1}{n} \xvec{1}^\prime (\xmat{I} - \lambda \xmat{W}_2^*)^{-1} \xmat{W}_1^* Cov( \xvec{\Delta} )  \xmat{W}_1^{* \prime} (\xmat{I} - \lambda \xmat{W}_2^{* \prime})^{-1} \xvec{1}  \rightarrow 0
			\end{equation}
			as $n$ tends to infinity then the condition (\ref{limit7}) is fulfilled.
		\end{itemize}
	\end{lemma}
	
	Note that the assumption (\ref{limit7}) is a statement about the topological structure of the underlying space. Moreover, (\ref{limit8}) shows that it can be interpreted as an assumption on the underlying autocorrelation structure of the process. It is fulfilled if $\xmat{W}_1^*$ and $\xmat{W}_2^*$ are sparse to limit the spatial dependence to a manageable degree. Here the choice of the weight matrices is restricted. It is also satisfied if the autocorrelation is weak.

	\section{Computational Implementation and Simulation Studies}
	
	Here, we assume the simple parametric setting given by \eqref{weight}. {We simulated a spatial GARCH process as specified in Example \ref{example:spGARCH} and the} weighting matrices $\xmat{W}^*_1$ and $\xmat{W}^*_2$ were set as row-standardized Rook contiguity matrices, for which the upper diagonal elements were set to zero to avoid negative values of $h(\xvec{s}_i)$. Thus, {Theorem \ref{solution} is fulfilled. In practice, such processes are relevant to model directional processes, for instance.}
	
	The simulation study is performed on a $d \times d$ spatial unit grid (i.e., $D_{\xvec{s}} = \{\xvec{s} = (s_1, s_2)' \in \xset{Z}^2 : 1 \leq s_1, s_2 \leq d \}$), resulting in $n = d^2$ observations, with $m = 10000$ replications. {The size of this spatial field has been successively increased with $d \in \{5, 10, 15\}$. Moreover, we have considered different settings depending on the data-generating parameters $\xvec{\theta_0} = (\rho_0, \lambda_0, \alpha_0)'$. Whereas the unconditional variance parameter $\alpha_0$ equals 1 for all settings (i.e., variance level which is independent of the spatial location), $\rho_0$ and $\lambda_0$ varied across the settings. To be precise, $\rho_0 \in \{0.2, 0.5, 0.8\}$ and $\lambda_0 \in \{0.2, 0.5, 0.8\}$ to have settings with a weak, moderate and large dependence in the conditional spatial heteroscedasticity.}
	
	{For all three parameters $\rho$, $\lambda$, and $\alpha$, the average RMSE is shown Table \ref{table:RMSE_full}. In general, our theoretical results are confirmed. That is, the parameters can correctly be estimated and the RMSE is decreasing with an increasing number of observations.} {Moreover, the non-linear least squares approach can efficiently be implemented and runs very fast on standard computers, even if the number of observations is large. On average, the computing time for estimating the parameters  ranges from 3.4 to 3.8 seconds for 25 observations, 3.2 - 4.8 seconds for 100 observations, and 5.5 - 8.5 seconds for 225 observations on a standard notebook. This method is implemented in the R package \texttt{spGARCH} (version $> 2.0$).}

\begin{table}
	\caption{RMSE of the estimates $\hat{\rho},\hat{\lambda},\hat{\alpha}$ for different settings with $\alpha_0 = 1$.}\label{table:RMSE_full}
	{\scriptsize
		\begin{center}
				\begin{tabular}{ccccccccccc}
				\hline
				&   & \multicolumn{3}{c}{$\rho_0 = 0.2$} & \multicolumn{3}{c}{$\rho_0 = 0.5$} & \multicolumn{3}{c}{$\rho_0 = 0.8$} \\
				&   & $\hat{\rho}$ & $\hat{\lambda}$ & $\hat{\alpha}$ & $\hat{\rho}$ & $\hat{\lambda}$ & $\hat{\alpha}$ & $\hat{\rho}$ & $\hat{\lambda}$ & $\hat{\alpha}$ \\
				\hline
				& $d = 5$   & 0.3832 & 0.4863 & 0.7947 & 0.5239 & 0.5481 & 1.2027 & 0.7088 & 0.6263 & 1.8927   \\
				$\lambda_0 = 0.2$  & $d = 10$  & 0.2286 & 0.3324 & 0.5356 & 0.3232 & 0.3115 & 0.7080 & 0.4056 & 0.2940 & 1.0782   \\
				& $d = 15$  & 0.1671 & 0.2937 & 0.4683 & 0.2241 & 0.2280 & 0.5356 & 0.2619 & 0.1735 & 0.7474   \\[.2cm]
				& $d = 5$   & 0.4119 & 0.5118 & 1.2540 & 0.5739 & 0.5841 & 1.9742 & 0.7753 & 0.6808 & 2.7150   \\
				$\lambda_0 = 0.5$  & $d = 10$  & 0.2360 & 0.3587 & 0.9992 & 0.3507 & 0.3319 & 1.5654 & 0.4633 & 0.3433 & 2.1774   \\
				& $d = 15$  & 0.1699 & 0.3129 & 0.8963 & 0.2334 & 0.2218 & 1.2153 & 0.2915 & 0.1914 & 1.7010   \\[.2cm]
				& $d = 5$   & 0.4624 & 0.5629 & 1.8340 & 0.6399 & 0.6284 & 2.4491 & 0.8460 & 0.7388 & 2.8362   \\
				$\lambda_0 = 0.8$  & $d = 10$  & 0.2502 & 0.3066 & 1.5493 & 0.3852 & 0.3244 & 1.7279 & 0.5118 & 0.3808 & 2.2565   \\
				& $d = 15$  & 0.1669 & 0.1960 & 1.1550 & 0.2486 & 0.1962 & 1.2094 & 0.3244 & 0.2236 & 1.9009   \\
				\hline
			\end{tabular}
		\end{center}
    }
\end{table}

	\section{Real-World Application: Condominium Prices in Berlin}

	In markets that are constrained in space, one can typically expect to find locally varying risks. Typical examples of such markets are real estate and labour. For the former, the property prices are highly dependent upon the location of the real estate and prices in the surrounding areas. Similarly, for the latter, this market is also often constrained in space due to the limited mobility of labourers. 
	
\begin{figure}
		\centering
		\includegraphics[width=0.8\textwidth]{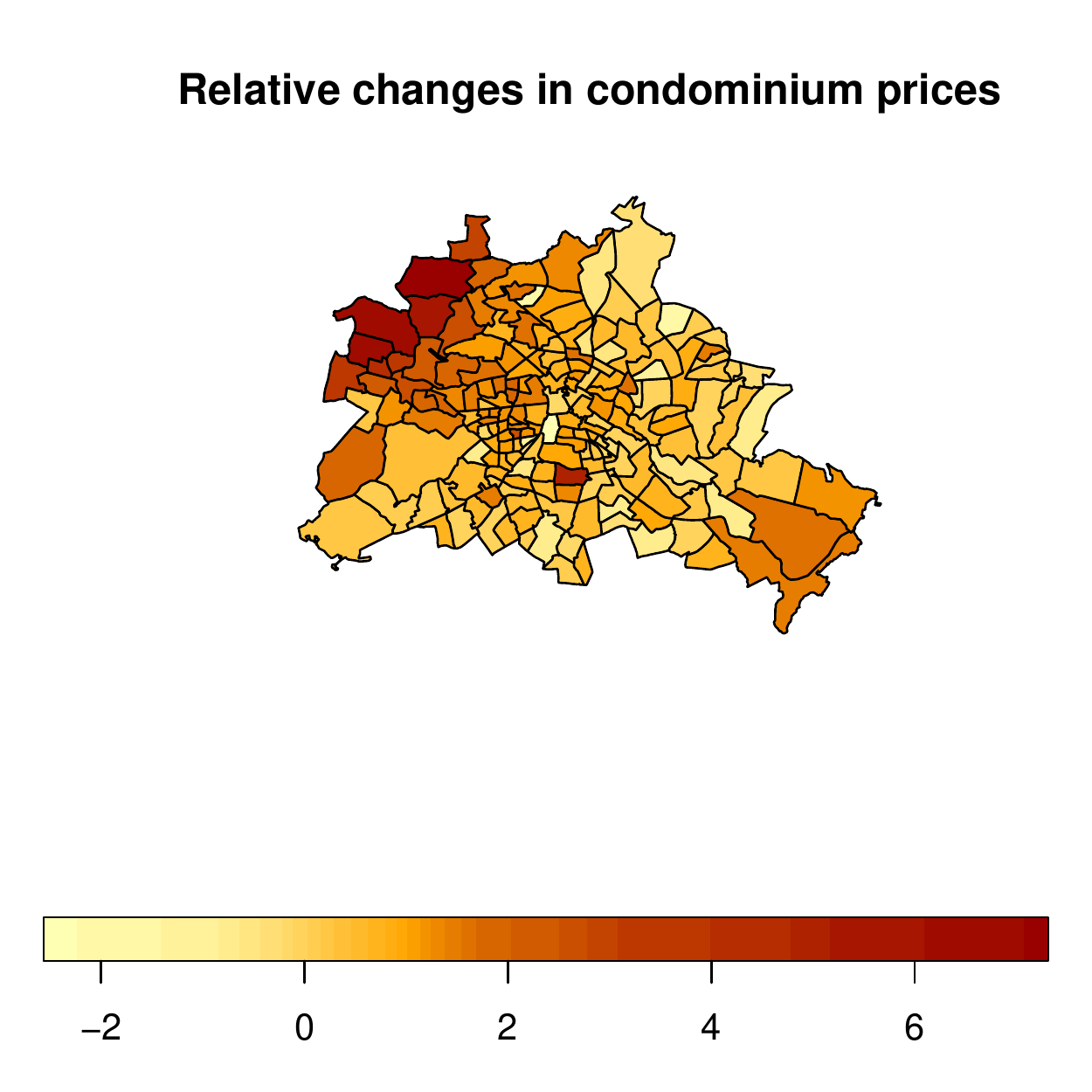}
		\caption{Relative changes (in per cent) of the Box-Cox transformed prices for all Berlin ZIP-Code regions.}\label{fig:example1} 
	\end{figure}

	On the one hand, we observe conditional mean levels that vary in space, so-called spatial clusters. That is, both clustered areas of higher prices and lower prices can be observed. On the other hand, we may also expect to find locally varying risks relating to price, which can be considered as local volatility clusters. The proposed spatial GARCH-type models are capable of capturing such spatial dependencies in the conditional variance. This motivates why we consider condominium prices at a fine spatial scale. In particular, we will analyze the relative changes in Box-Cox transformed prices from 1995--2014 across all Berlin ZIP-Code regions (i.e., $n = 190$). The data are depicted in Figure \ref{fig:example1}. The sample mean for these price changes is 0.8103 with a median of 0.6965. In total, the price changes range from -2.5650 to 7.3131. We can observe a spatial cluster of positive values in the north-western ZIP-Code regions.
	
	{However, this fine spatial scale of all ZIP-Code regions causes another problem, namely exogenous regressors are often not available or cannot be assigned in the given small-scale resolution. For instance, the average household income could play an important role in the increase of the condominium prices, but the place of living (in terms of ZIP-Code areas) does not usually coincide with the place of work. There is no reliable way to associate quantities like personal or household income with ZIP-Code areas. Moreover, local infrastructure like schools, leisure facilities, parks is not limited to the residents of the respective ZIP-code areas. Thus, modelling an empirical process on such a small spatial scale is typically prone to heteroscedasticity induced by latent variables.}
	
	{To illustrate these effects, we applied the developed spatial GARCH model to the residuals of a spatial autoregressive model, briefly SAR (see, e.g. \citealt{Halleck15,Lee04}), with and without exogenous regressors. More precisely, we select the regressors from a set of potential covariates available for different spatial scales, including the number of crimes (in so-called life-world oriented spaces, LOR, 161 units), number of schools, kinder gardens (ZIP-code level), percentage of migrants (LOR level), number of inhabitants (LOR level), size of areas used for infrastructure, living, water, vegetation (district level, 12 units), and average net income per household (district level). The included regressors were chosen such that the Akaike information criterion is minimal. The results of these two models are shown in Table \ref{table:results_ex1} along with the spGARCH coefficients of the error process. Regarding the residuals of these two mean models with and without regressors (before fitting a spGARCH model to the residuals), we observe that both of them are not autocorrelated in space (Moran's $I = -0.0375$ with a p-value of $0.7811$ for the intercept-only model, and $I = -0.0124$ with $p = 0.5676$ for the model including covariate effects). That is, the spatial correlation of original data could fully be modelled ($I = 0.5104$). However, looking at the absolute values of the residuals, we observe that there is significant autocorrelation for both models ($I = 0.1027$ ($p = 0.0042$) and $I = 0.0808$ ($p = 0.0181$) for the intercept-only model and regressive model, respectively).} 
	
	{In the final regressive model, only four regressors were selected, namely a proxy for the available free space (i.e., the proportion of settlement area to total area), the net household income in each district, and linear trends in the east-west direction, and north-south direction have been included (i.e., the coordinates of the centroids of each ZIP-code unit). While a significant negative trend can be observed from west to east, the increase in the north-south direction seems to be of minor importance. Furthermore, the average household income clearly influences the price development of condominiums in Berlin. Condominiums in high-class districts in terms of average income have increased relatively less than condominiums in lower-income areas.}
	
	{It is worth noting that the dependence in the conditional heteroscedasticity could partly be covered by these covariates and the spatial autocorrelation in the absolute residuals was reduced. Nevertheless, latent effects are present which were not modelled by including these regressors. Thus, an spGARCH model has been fitted in a second step to the residuals of both models. The obtained spGARCH parameters can be interpreted as local model uncertainties and simultaneously cover latent variables, which could not be included due to the fine spatial scale of ZIP-Code levels.}
	
	\begin{table}
		\caption{Estimated parameters of the spGARCH model for the residuals of a spatial autoregressive model with and without regressors, where the dependent variables are the changes in the condominium prices in Berlin. All standard errors are given in parentheses.}\label{table:results_ex1}
		\centering
		
		{\scriptsize
			\begin{tabular}{l cc cc}
				& \multicolumn{2}{c}{Without regressors}  & \multicolumn{2}{c}{With regressors}        \\
				Parameter                                 &  Estimate & Standard error           & Estimate & Standard error          \\
				\hline
				\emph{Regression coefficients}                           &              &                              &               &                             \\
				$\quad$  Intercept                                                                                                           &   0.2002  &  (0.0814)        &     -53.8461          &     (70.1343)                        \\
				$\quad$  Available area                                                            &      \multicolumn{2}{c}{-}          &        -0.0282        &   (0.0151)                          \\ 
				$\quad$  Net income                                                                                                       &     \multicolumn{2}{c}{-}           &        -0.0014         &    (0.0006)                         \\
				$\quad$  West-east gradient                                                                                          &      \multicolumn{2}{c}{-}           &        -2.3621       &     (0.7337)                        \\ 
				$\quad$  North-south gradient                                                                                       &      \multicolumn{2}{c}{-}           &        1.6939       &       (1.3219)                      \\ 
				\emph{SAR process}                                            &              &                              &               &                             \\
				$\quad$ $\gamma$                                              &  0.7456             &    (0.0588)                          &   0.5758             &   (0.0789)                           \\
				\emph{spGARCH residuals}                                 &              &                              &               &                                \\
				$\quad$ $\alpha$                                                & 0.0020    & (0.0122)                & 0.0000   & (0.0123)                               \\
				$\quad$ $\rho$                                                   & 0.2136   & (0.0419)                & 0.2022   & (0.0390)                               \\
				$\quad$ $\lambda$                                             & 0.7091   & (0.0566)                & 0.7016   & (0.0578)                                \\
				\emph{Summary statistics}                                 &          &                         &          &                                        \\
				$\quad$ AIC of the mean model                                & \multicolumn{2}{c}{539.9281}       & \multicolumn{2}{c}{532.543}                \\
				$\quad$ $I$ res. ($p$-value)                 &  \multicolumn{2}{c}{0.0806  (0.9623)}                &  \multicolumn{2}{c}{-0.0376 (0.7774)}                           \\
				$\quad$ $I$ squ. res. ($p$-value)         &  \multicolumn{2}{c}{-0.0820  (0.9721)}                &   \multicolumn{2}{c}{-0.0900 (0.9821)}                           \\
				$\quad$ Average ${\xvec{h}}$        &  \multicolumn{2}{c}{0.6039}               &   \multicolumn{2}{c}{0.5404}                          \\
				$\quad$ Max. ${\xvec{h}}$            &  \multicolumn{2}{c}{4.8183}               &   \multicolumn{2}{c}{4.0335}                          \\
			\end{tabular}
		}
	\end{table}

	{In both cases, we observe significant positive dependence in the conditional second moments. More precisely, the GARCH effects amount to $\hat{\rho} = 0.2136$ and $\hat{\rho} = 0.2022$ with $\hat{\lambda}$ roughly equal to 0.70 for the intercept-only and regression model, respectively. These parameters can similarly be interpreted as in the time-series case, although $\xvec{h}$ does not necessarily coincide with the conditional second moments (see \citealt{Otto19_statpapers}). Analyzing the residuals of this combined model shows that the remaining dependence in the heteroscedasticity could be explained. The squared residuals are no longer significantly autocorrelated in space.}
	
\begin{figure}
		\centering
		\includegraphics[width=0.9\textwidth]{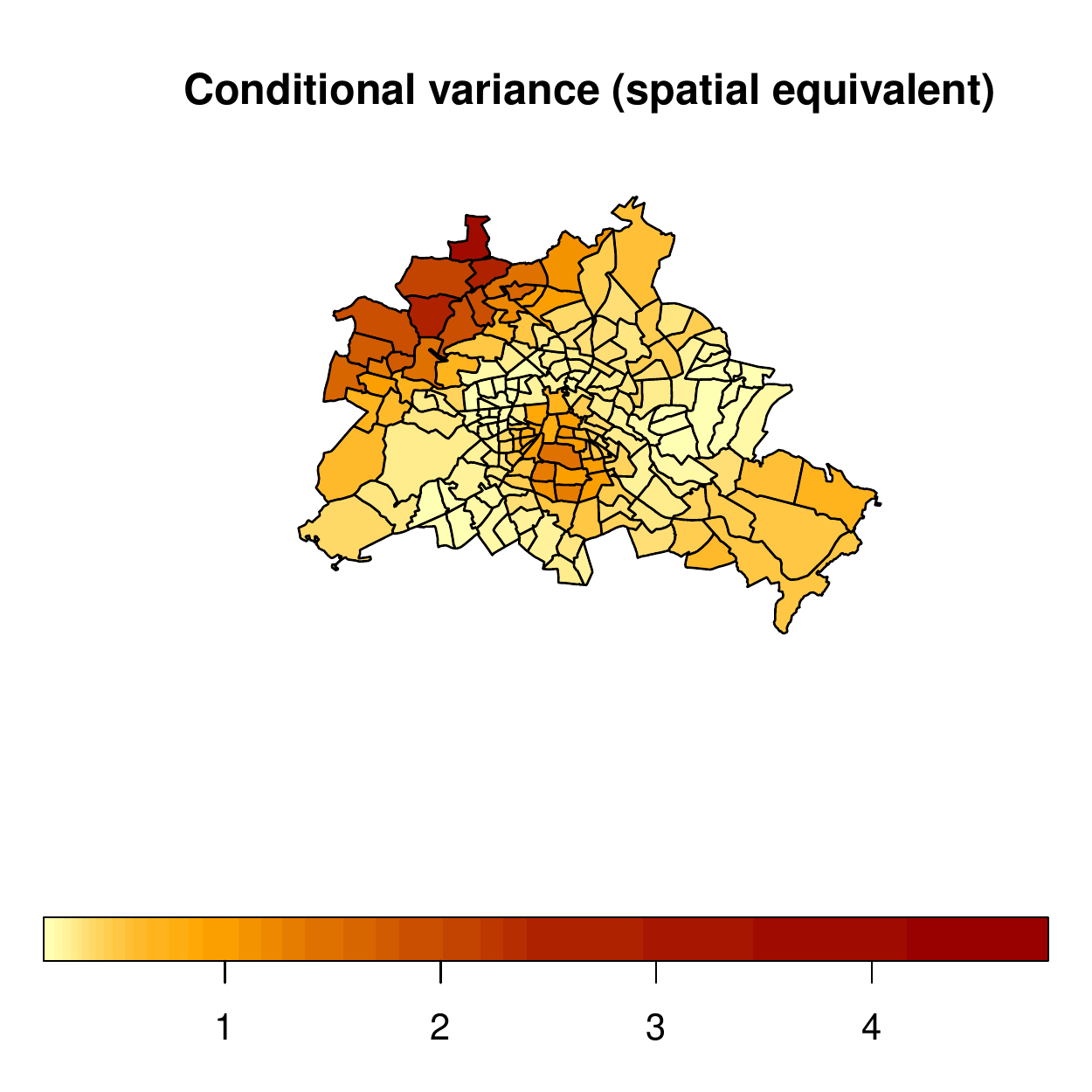}
		\caption{Spatial equivalent of the conditional variance.}\label{fig:condvar}
	\end{figure}

	{The resulting conditional variance is visualized for the model with regressors in Figure \ref{fig:condvar}. The highest values of $h(\xvec{s}_i)$ can be observed for the outer regions in north-west and another cluster is located in the southern city centre. This indicates that the highest uncertainty in the price changes is observed for these regions, while there is a band around the city centre where the price changes could more accurately be predicted by the regression model (i.e., the estimated values of $h(\xvec{s}_i)$ are lower). These results seem to be very reasonable because the real-estate market was changing the most in these areas. First, due to increasing prices in the centre, new land for building has been created outside the city -- mostly along the regional transport tracks which are mainly going in an east-west direction. Second, the major airport of West Berlin, namely the airport Berlin-Tempelhof, was closed in 2008 changing the atmosphere in this region, some ZIP-code areas changed from regions in the flight paths to calm regions very close to the city centre, while others were not affected. This explains the second cluster in the centre.} 
	
	\section{Discussion and Conclusions}\label{sec:conclusion}

	Recently, a few papers have introduced spatial ARCH and GARCH-type models that allow the modelling of an instantaneous spatial autoregressive dependence of heteroscedasticity. In this paper, we propose a {generalized spatial ARCH model} that {additionally} covers all previous approaches. Due to the flexible definition of the model as a set of functions, we can derive a common estimation strategy for all these spatial GARCH-type models. {It} is based on non-linear least squares.
	
	{In the second part of the paper, we confirmed our theoretical findings on the consistency of the estimators by means of Monte Carlo simulation studies. The estimation method is computationally implemented in the R package \texttt{spGARCH}.} Eventually, the use of the model was demonstrated through an empirical example. More precisely, this paper has shown how the model uncertainties of local price changes in the real estate market in Berlin can be described using an spGARCH model {as residuals' process}. Though all proposed models are uncorrelated and have a zero mean, potential interactions between the error process and the mean equation should be analyzed in greater detail in future research.

\begin{figure}
		\centering
		\includegraphics[width=0.4\textwidth]{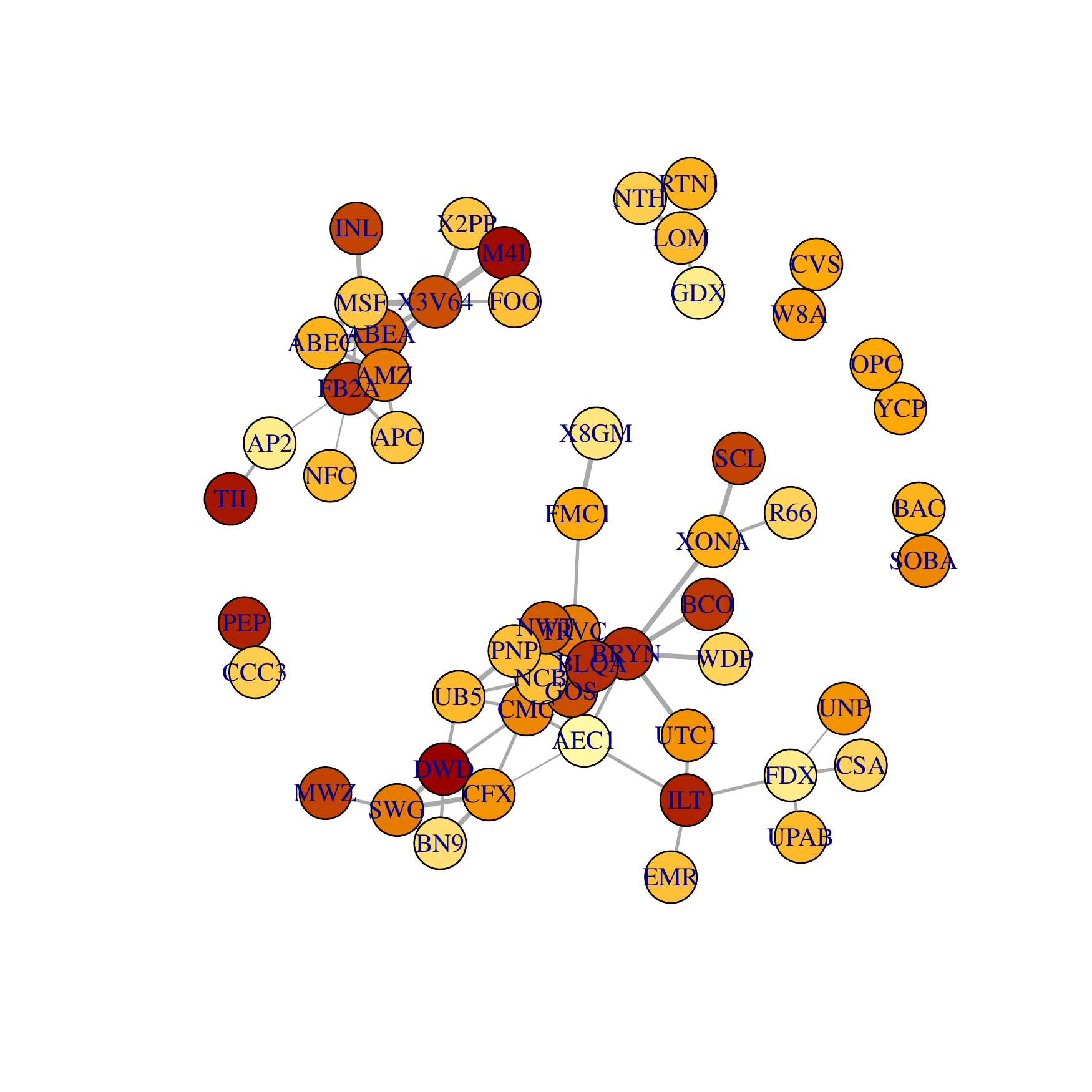}
		\caption{Financial network of selected stocks of the S\&P 500, where the colour of the nodes denotes the annual returns in 2017 with darker colours indicating higher returns.}\label{fig:example2}
	\end{figure}

	In addition, we want to stress that the dependence structure does not necessarily have to be interpreted in a spatial sense. Thus, we briefly discuss a further example below, on which the ``spatial'' proximity could also be defined as the edges of networks. In such cases,  $\xmat{W}_1$ and $\xmat{W}_2$ would be interpreted as adjacency matrices. For instance, one might consider the financial returns of several stocks as a network, where the only assets that are connected are those that are correlated above a certain threshold. This could create a financial network, as shown in Figure \ref{fig:example2}. Thus, spGARCH models can be used to analyze various forms of information, whether that might be volatility, risk, or spill-overs from one stock to another, if these assets are close to one another within a certain network. In future research, attempts for modelling volatility clusters within networks, using spatial GARCH models, should be analyzed in greater detail.
	
	{Up to now}, we have assumed that suitable functions of the {spGARCH} model framework are known. Hence, it is possible to maximize certain goodness-of-fit criteria in order to obtain the best-fitting model. However, these functions can also be estimated using a non-parametric approach; for instance by penalized or classical B-splines. {Besides, further choices of $f$ have not been discussed in this paper yet, including choices of $f$ to obtain E-spGARCH or logarithmic spGARCH models. Also, multivariate models remain open for future research. This}  will be the subject of {some} forthcoming papers.
	
	\section{Appendix}
	\section*{Proofs}

	
	\begin{proof}[Theorem \ref{solution}]
		Inserting \eqref{eq:initial} into \eqref{eq:unified}, we get that
		\[ \xvec{F}(\xvec{h}) = \xvec{\alpha} + \xmat{W}_1 \xvec{\gamma}(\xmat{E} \xvec{h}) - (\xmat{I} - \xmat{W}_2) \xvec{F}(\xvec{h}) = \xvec{0} . \]
		Now, we want to know whether there is a solution $\xvec{h}$ of this equation. {This is equivalent to the problem of whether or not $\xvec{T}(\omega) \circ \xvec{h}$ has a
			fixed-point. The existence and the uniqueness of a solution $\xvec{h}$ is an immediate consequence of the Banach fixed-point theorem. Here we make use of the randomized version given as Theorem 7 of \cite{bharucha1976fixed}. Inserting this solution in \eqref{eq:initial} provides a measurable solution $\xvec{Y}$.}
	\end{proof}

	\begin{proof}[Corollary \ref{cor:stationarity2}]
		Because $(Y(\xvec{s}_1), \ldots, Y(\xvec{s}_n))^\prime$ is a measurable function of \linebreak $(\varepsilon(\xvec{s}_1), \ldots, \varepsilon(\xvec{s}_n))^\prime$, it is strictly stationary as well.
	\end{proof}

	\begin{proof}[Theorem \ref{th:moments1}]
		\begin{itemize}
			\item[a)] Since
			\[ \xvec{F} = ( f(\xvec{h}(\xvec{s}_i)) )_{i=1,...,n} = \xvec{\alpha} + \xmat{W}_1 \xvec{\gamma}(\xmat{E} \xvec{h}) + \xmat{W}_2 \xvec{F} \]
			its solution $\xvec{h}$ is a function of $\varepsilon(\xvec{s}_1)^2,..., \varepsilon(\xvec{s}_n)^2$, say $h(\xvec{s}_i) = \xi_i(\varepsilon(\xvec{s}_1)^2,..., \varepsilon(\xvec{s}_n)^2)$. Since $Y(\xvec{s}_i) = \varepsilon(\xvec{s}_i) \sqrt{h(\xvec{s}_i)}$ it follows that
			\begin{eqnarray*}
				-Y(\xvec{s}_i) & = & -\varepsilon(\xvec{s}_i) \sqrt{\xi_i(\varepsilon(\xvec{s}_1)^2,..., \varepsilon(\xvec{s}_n)^2)} \\
				& = & -\varepsilon(\xvec{s}_i) \sqrt{\xi_i(\varepsilon(\xvec{s}_1)^2,..., (-\varepsilon(\xvec{s}_i))^2,...,\varepsilon(\xvec{s}_n)^2)} \\
				& \stackrel{d}{=} & Y(\xvec{s}_i)
			\end{eqnarray*}
			since $\xvec{\varepsilon}$ is sign-symmetric. Thus $Y(\xvec{s}_i)$ is a symmetric random variable.
			
			Moreover,
			\[ (Y(\xvec{s}_1),\ldots,Y(\xvec{s}_n))^\prime \stackrel{d}{=} (-Y(\xvec{s}_1),\ldots,Y(\xvec{s}_n))^\prime. \]
			Thus, $E(Y(\xvec{s}_1)^{2k-1} | Y(\xvec{s}_2),\ldots,Y(\xvec{s}_n)) = E(-Y(\xvec{s}_1)^{2k-1} | Y(\xvec{s}_2),\ldots,Y(\xvec{s}_n))$. Consequently, this quantity is zero.
			
			\item[b)] Now,
			\begin{eqnarray*}
				Y(\xvec{s}_i) Y(\xvec{s}_j) & = & \varepsilon(\xvec{s}_i) \varepsilon(\xvec{s}_j) \sqrt{\xi_i(\varepsilon(\xvec{s}_1)^2,..., \varepsilon(\xvec{s}_n)^2)}
				\sqrt{\xi_j(\varepsilon(\xvec{s}_1)^2,..., \varepsilon(\xvec{s}_n)^2)}\\
				& \stackrel{d}{=} & - \varepsilon(\xvec{s}_i) \varepsilon(\xvec{s}_j) \sqrt{\xi_i(\varepsilon(\xvec{s}_1)^2,..., \varepsilon(\xvec{s}_n)^2)}
				\sqrt{\xi_j(\varepsilon(\xvec{s}_1)^2,..., \varepsilon(\xvec{s}_n)^2)}\\
				& = & - Y(\xvec{s}_i) Y(\xvec{s}_j)
			\end{eqnarray*}
			and thus $Cov(Y(\xvec{s}_i), Y(\xvec{s}_j)) = 0$ for $i \neq j$.
			
		\end{itemize}
	\end{proof}

	\begin{proof}[{Proof of Theorem \ref{th:consistency}}]
		
		{To} prove the above theorem we make use of \cite{Newey94}. First we observe that $\Theta$ is a compact set. Moreover, $Q_n(\xvec\vartheta)$ (see (\ref{min})) is a continuous function for $\xvec\vartheta \in \Theta$. Now we have to show that
		\begin{itemize}
			
			\item[(i)] $Q_n(\xvec\vartheta) \stackrel{p}{\rightarrow} Q_0(\xvec\vartheta)$ for all $\xvec\vartheta \in \Theta$
			
			\item[(ii)]  there is $\tau > 0$ and $\hat{B}_n = O_p(1)$ such that
			\[ | Q_n(\tilde{\xvec\vartheta}) - Q_n( \xvec\vartheta) | \le \hat{B}_n || \tilde{\xvec\vartheta} - \xvec\vartheta ||^\tau  \quad \forall \tilde{\xvec\vartheta}, \xvec\vartheta \in \Theta  . \]
		\end{itemize}
		
		If (i) and (ii) hold then it follows with Lemma 2.9 of \cite{Newey94} that $sup_{\xvec\vartheta \in \Theta} |Q_n(\xvec\vartheta) - Q_0(\xvec\vartheta) | \stackrel{p}{\rightarrow} 0$ as $n \rightarrow \infty$.
		
		If additionally
		\begin{itemize}
			\item[(iii)] $Q_0(\xvec\vartheta)$ is continuous for $\xvec\vartheta \in \Theta$
		\end{itemize}
		holds, then we get with Lemma 3 of \cite{Amemiya85} that $\hat{\xvec\vartheta}_n \stackrel{p}{\rightarrow} \xvec\vartheta_0$ as $n \rightarrow \infty$ and the result is proved.
		
		We start with proving (i). Recall that $H_i = \log(h_{\xvec{\vartheta}_0}(\xvec{s}_i)) + \log(\varepsilon(\xvec{s}_i)^2) - E(\log(\varepsilon(\xvec{s}_i)^2)) = \eta(\xvec{s}_i) + \log(h_{\xvec{\vartheta}_0}(\xvec{s}_i))$. Thus, in the present case, the function $Q_0(\xvec{\vartheta})$ can be found by splitting $Q_n(\xvec{\vartheta})$ into eight parts, as follows
		\[ Q_n(\xvec{\vartheta}) = I_n + II_n + III_n + IV_n + V_n + VI_n + VII_n + VIII_n \]
		with
		\begin{small}
			\begin{eqnarray*}
				I_n    & = & \frac{1}{n} \sum_{i=1}^n \eta(\xvec{s}_i)^2, \quad II_n = \frac{1}{n} \sum_{i=1}^n \left( E( \mbox{log}(h_{\xvec{\vartheta}}(\xvec{s}_i)) - E(\mbox{log}(h_{\xvec{\vartheta}_0}(\xvec{s}_i)) ) \right)^2 , \\
				III_n  & = & \frac{1}{n} \sum_{i=1}^n \left( \mbox{log}(h_{\xvec{\vartheta}_0}(\xvec{s}_i)) - E( \mbox{log}(h_{\xvec{\vartheta}_0}(\xvec{s}_i))) \right)^2 \\
				IV_n   & = & \frac{1}{n} \sum_{i=1}^n \left( \mbox{log}(h_{\xvec{\vartheta}}(\xvec{s}_i)) - E( \mbox{log}(h_{\xvec{\vartheta}}(\xvec{s}_i))) \right)^2 \\
				V_n    & = & \frac{2}{n} \sum_{i=1}^n \eta(\xvec{s}_i)  \left( \mbox{log}(h_{\xvec{\vartheta}_0}(\xvec{s}_i)) - \mbox{log}(h_{\xvec{\vartheta}}(\xvec{s}_i) ) \right)  \\
				VI_n   & = & \frac{2}{n} \sum_{i=1}^n (E(\log(h_{\xvec{\vartheta}_0}(\xvec{s}_i))) - E(\log(h_{\xvec{\vartheta}}(\xvec{s}_i)))) (\log(h_{\xvec{\vartheta}_0}(\xvec{s}_i)) - E(\log h_{\xvec{\vartheta}_0}(\xvec{s}_i))) ,  \\
				VII_n  & = & -\frac{2}{n} \sum_{i=1}^n (E(\log(h_{\xvec{\vartheta}_0}(\xvec{s}_i))) - E(\log(h_{\xvec{\vartheta}}(\xvec{s}_i)))) (\log(h_{\xvec{\vartheta}}(\xvec{s}_i)) - E(\log h_{\xvec{\vartheta}}(\xvec{s}_i))) ,  \\
				VIII_n & = & -\frac{2}{n} \sum_{i=1}^n (\log(h_{\xvec{\vartheta}_0}(\xvec{s}_i)) - E(\log h_{\xvec{\vartheta}_0}(\xvec{s}_i))) (\log(h_{\xvec{\vartheta}}(\xvec{s}_i)) - E(\log h_{\xvec{\vartheta}}(\xvec{s}_i))).  \\
			\end{eqnarray*}
		\end{small}

		The weak law of large numbers (WLLN) implies that $I_n$ converges in probability to $\mbox{Var}( \mbox{log}(\varepsilon(\xvec{s}_1)^2) )$.
		$III_n$ is a special case of $IV_n$. Further
		\begin{eqnarray*}
			V_n/2 & =  & \frac{1}{n} \sum_{i=1}^n \eta(\xvec{s}_i)  \left( \log(h_{\xvec{\vartheta}_0}(\xvec{s}_i)) - \log(h_{\xvec{\vartheta}}(\xvec{s}_i) ) \right)\\
			& = &  \frac{1}{n} \sum_{i=1}^n \eta(\xvec{s}_i)  \left( E\left( \log(h_{\xvec{\vartheta}_0}(\xvec{s}_i))\right) - E(\left(\log(h_{\xvec{\vartheta}}(\xvec{s}_i) ) \right)  \right)\\
			& & +  \frac{1}{n} \sum_{i=1}^n \eta(\xvec{s}_i)  \left( E\left( \log(h_{\xvec{\vartheta}}(\xvec{s}_i)) \right) - \log(h_{\xvec{\vartheta}}(\xvec{s}_i) ) \right)\\
			& & +  \frac{1}{n} \sum_{i=1}^n \eta(\xvec{s}_i)  \left( \log(h_{\xvec{\vartheta}_0}(\xvec{s}_i)) - E\left( \log(h_{\xvec{\vartheta}_0}(\xvec{s}_i) ) \right) \right) \\
			& = & V_{n,1} + V_{n,2} + V_{n,3} .
		\end{eqnarray*}
		
		Since $\{ \eta(\xvec{s}_i) \}$ are independent  and identically distributed, it follows that
		\[ P( | V_{n,1} | > \varepsilon ) \le \frac{E(\eta(\xvec{s}_1)^2)}{n^2 \varepsilon^2} \; \sum_{i=1}^n \left( E( \mbox{log}(h_{\xvec{\vartheta}}(\xvec{s}_i)) - E(\mbox{log}(h_{\xvec{\vartheta}_0}(\xvec{s}_i)) ) \right)^2 \rightarrow 0\]
		as $n$ tends to $\infty$ and thus $V_{n,1}$ converges in probability to zero. Using the Cauchy-Schwarz inequality and (\ref{limit2}) we get that $V_{n,2}$ and $V_{n,3}$ converge in probability to zero. Thus $V_n \stackrel{p}{\rightarrow} 0$ as $n \rightarrow \infty$.
		
		Moreover, the mixed quantities $VI_n$, $VII_n$, and $VIII_n$ convergence to zero in probability. This is obtained by applying the Cauchy-Schwarz inequality and by making use of the previous results and assumptions.
		
		Consequently, if $\tilde{Q}_0(\xvec{\vartheta})$ denotes the limit in (\ref{limit1}), then $Q_0(\xvec{\vartheta}) = \mbox{Var}( \log(\varepsilon(\xvec{s}_1)^2)) +
		\tilde{Q}_0(\xvec{\vartheta})$ and part i) is proved.\\
		
		Next, we prove part (ii). We use that $\tau^\prime(\tau^{-1}(x)) = f^\prime(f^{-1}(x)) f^{-1}(x)$ and, thus,
		$\tau^\prime(\tau^{-1}(\alpha c_i(\lambda) + \rho \xvec{d}_i(\lambda) (\tilde{\gamma}(H_v+c)))_{v=1,...,n} ) = f^\prime(h_{\xvec{\vartheta}}(\xvec{s}_i)) h_{\xvec{\vartheta}}(\xvec{s}_i)$. Consequently,
		\[ Q_n(\tilde{\xvec{\vartheta}}) - Q_n(\xvec{\vartheta})  =  ( \tilde{\xvec{\vartheta}} - \xvec{\vartheta})^\prime Q_n^\prime( \xvec{\vartheta}^* ) \]
		with

		\[ Q_n^\prime( \xvec{\vartheta}^* ) = \left( \begin{array}{l}  -\frac{2}{n} \sum_{i=1}^n \frac{ H_i - \log(h_{\xvec{\vartheta}^*}(\xvec{s}_i)) }{f^\prime(h_{\xvec{\vartheta}^*}(\xvec{s}_i)) \, h_{\xvec{\vartheta}^*}(\xvec{s}_i)}
			\; \xvec{d}_i(\lambda^*)^\prime \left(\tilde{\gamma}( H_v + c ) \right) \\
			-\frac{2}{n} \sum_{i=1}^n \frac{ H_i - \log(h_{\xvec{\vartheta}^*}(\xvec{s}_i)) }{f^\prime(h_{\xvec{\vartheta}^*}(\xvec{s}_i)) \, h_{\xvec{\vartheta}^*}(\xvec{s}_i)}
			\; \left( \alpha^* c_i^\prime(\lambda^*) + \rho^* \xvec{d}_i^\prime(\lambda^*)^\prime \left( \tilde{\gamma}( H_v + c )\right) \right)   \\
			-\frac{2}{n} \sum_{i=1}^n \frac{ H_i - \log(h_{\xvec{\vartheta}^*}(\xvec{s}_i)) }{f^\prime( h_{\xvec{\vartheta}^*}(\xvec{s}_i) ) \,  h_{\xvec{\vartheta}^*}(\xvec{s}_i)}
			\; c_i(\lambda^*)
		\end{array} \right) . \]

		Since $H_i = \eta(\xvec{s}_i) + \log(h_{\xvec{\vartheta}_0}(\xvec{s}_i)$, we get that
		\[ Q_n^\prime( \xvec{\vartheta}^* ) = \left( \begin{array}{l}  -\frac{2}{n} \sum_{i=1}^n \frac{  \eta(\xvec{s}_i) + \log(h_{\xvec{\vartheta}_0}(\xvec{s}_i) - \log(h_{\xvec{\vartheta}^*}(\xvec{s}_i)) }{f^\prime(h_{\xvec{\vartheta}^*}(\xvec{s}_i)) \, h_{\xvec{\vartheta}^*}(\xvec{s}_i)}
			\; \xvec{d}_i(\lambda^*)^\prime \left(\gamma(  Y(\xvec{s}_v )^2 ) \right) \\
			-\frac{2}{n} \sum_{i=1}^n \frac{  \eta(\xvec{s}_i) + \log(h_{\xvec{\vartheta}_0}(\xvec{s}_i) - \log(h_{\xvec{\vartheta}^*}(\xvec{s}_i)) }{f^\prime(h_{\xvec{\vartheta}^*}(\xvec{s}_i)) \, h_{\xvec{\vartheta}^*}(\xvec{s}_i)}
			\; \left( \alpha^* c_i^\prime(\lambda^*) + \rho^* \xvec{d}_i^\prime(\lambda^*)^\prime \left( \gamma( Y(\xvec{s}_v )^2 )\right) \right)   \\
			-\frac{2}{n} \sum_{i=1}^n \frac{  \eta(\xvec{s}_i) + \log(h_{\xvec{\vartheta}_0}(\xvec{s}_i) - \log(h_{\xvec{\vartheta}^*}(\xvec{s}_i)) }{f^\prime(h_{\xvec{\vartheta}^*}(\xvec{s}_i)) \, h_{\xvec{\vartheta}^*}(\xvec{s}_i)}
			\; c_i(\lambda^*)
		\end{array} \right) . \]
		
		Note that
		\[ \left( c_i(\lambda) \right) = \left( \xmat{I} + \lambda \xmat{W}_2^* + \lambda^2 {\xmat{W}_2^*}^2 + ...\right) \xvec{1} = \frac{1}{1-\lambda} \xvec{1} . \]
		This quantity is well defined since $||\lambda \xmat{W}_2^*||_{\infty} = \lambda < 1$. Thus, $c_i(\lambda_l) \le c(\lambda) \le c(\lambda_u)$. Further,
		\[ \xvec{d}_i(\lambda)^\prime = \left( \xmat{I} + \lambda \xmat{W}_2^* + \lambda^2 {\xmat{W}_2^*}^2 + ...\right) \xmat{W}_1^* \]
		is a non-decreasing function in $\lambda$ for each component of the matrix and  $\xvec{d}_i(\lambda)^\prime \xvec{1} = 1/(1-\lambda)$.

		We obtain with the inequality of Cauchy-Schwarz that
		\[ |Q_n(\tilde{\xvec{\vartheta}}) - Q_n(\xvec{\vartheta})|  \le   || \tilde{\xvec{\vartheta}} - \xvec{\vartheta})||_2 || Q_n^\prime( \xvec{\vartheta}^* )  ||_2 \]
		where $||\cdot||_2$ stands for the Euclidean distance. Each component of $Q_n^\prime( \xvec{\vartheta}^* )$ is stochastically bounded. For the first component, this can be seen by applying Cauchy-Schwarz and making use of the fact that
		\[   \max_{\lambda \in [\lambda_l, \lambda_u]} \frac{1}{n} \sum_{i=1}^n   \left( \xvec{d}_i(\lambda)^\prime \left(\gamma(  Y(\xvec{s}_v )^2 ) \right)  \right)^2 \le  \frac{1}{n} \sum_{i=1}^n \sum_{j,v=1}^n d_{ij}(\lambda_u) d_{iv}(\lambda_u) |\gamma(Y(\xvec{s}_j)^2)| |\gamma(Y(\xvec{s}_v)^2)| \]
		and, thus,
		\[ P( |  \max_{\lambda \in [\lambda_l, \lambda_u]} \frac{1}{n} \sum_{i=1}^n   \left( \xvec{d}_i(\lambda)^\prime \left(\gamma(  Y(\xvec{s}_v )^2 ) \right) \right)^2 > K) \le \frac{E( (\gamma(  Y(\xvec{s}_1)^2  ))^2 )}{(1-\lambda_u)^2 K} . \]
		For the other components, the argumentation is similar. Thus, we have proved (ii).\\

		Finally, we come to part (iii). Let $\xvec{\vartheta}_1$, $\xvec{\vartheta}_2 \in \Theta$. Since by (ii)
		\[ | Q_n(\xvec{\vartheta}_2) - Q_n(\xvec{\vartheta}_1) | \le \hat{B}_n ||\xvec{\vartheta}_2 - \xvec{\vartheta}_1 ||_2 \]
		with $\hat{B}_n = O_p(1)$ and since because of the uniform convergence there is a subsequence such that $sup_{\xvec{\vartheta} \in \Theta} |Q_{r_n}(\xvec{\vartheta}) - Q_0(\xvec{\vartheta})| \stackrel{a.s.}{\rightarrow} 0$ as $n \rightarrow \infty$
		it follows with a constant $K$ that
		
		\[ |Q_0(\xvec{\vartheta}_2) - Q_0(\xvec{\vartheta}_1) | \le \lim_{n \to \infty} |Q_{r_n}(\xvec{\vartheta}_2) - Q_{r_n}(\xvec{\vartheta}_1)| \le K ||\xvec{\vartheta}_2 - \xvec{\vartheta}_1||_2 \]
		and thus $Q_0$ is continuous. Part (iii) is proved and the proof of the Theorem is finished.

	\end{proof}

	\begin{proof}[{Proof of Theorem \ref{th:consistency2}}]

		Following Theorem 4.1.2 in \cite{Amemiya85}, we have to prove that there is an open neighbourhood $N \subseteq \xvec{\Theta}$ of $\xvec{\vartheta}$ with
		\begin{itemize}
			\item[] $Q_n(\xvec{\vartheta}) \stackrel{p}{\rightarrow} Q_0(\xvec{\vartheta})$ as $n \rightarrow \infty$ uniformly in $\xvec{\vartheta} \in N$.
		\end{itemize}
		
		In order to prove this we follow the proof of {Theorem \ref{th:consistency}}. It follows under the modified assumptions as above that $Q_n(\xvec{\vartheta}) \stackrel{p}{\rightarrow} Q_0(\xvec{\vartheta})$ for $\xvec{\vartheta} \in N$ as $n \rightarrow \infty$. Using (ii) of the proof of {Theorem \ref{th:consistency},} we get that the convergence is uniform on a closed set $N_0 \subset N$. Thus it is also uniform on an open subset of $N_0$ since $\xvec{\Theta}$ is a subspace of the Euclidean space.
		
		The rest follows as in the proof of {Theorem \ref{th:consistency}}.

	\end{proof}

	\begin{proof}[{Proof of Lemma \ref{lemma:spGARCH}}]

		First we prove part a). Since $f(x) = x$ we get that  $\tau^{-1} = \mbox{log}(x)$ and
		\[ \log(h_{\xvec{\vartheta}}(\xvec{s}_i) =\log(\alpha c_i(\lambda) + \rho \xvec{d}_i(\lambda)^\prime \left( \gamma( Y(\xvec{s}_i)^2 ) \right) . \]
		It holds for $z > -1$ that
		
		\begin{equation}\label{log}
			\log(1+z) = z + \frac{z^2}{2} ( - \frac{1}{1+\zeta^2}) , \quad |\zeta| \le z
		\end{equation}
		and therefore $| \log(1+z) - z| \le z^2$.
		
		Let $Z_{in} =  \frac{\xvec{d}_i(\lambda)^\prime}{c_i(\lambda)} \left( \gamma( Y(\xvec{s}_i)^2 ) \right)$, $\kappa = \rho/\alpha$, and $\kappa_i = \kappa/(1+\kappa E(Z_{in}))$. Since $\{ \gamma( Y(\xvec{s}_i)^2 ) \}$ is strictly stationary as well it follows that $E(Z_{in}) = E(  \gamma( Y(\xvec{s}_1)^2 ) ) \; \xvec{d}_i(\lambda)^\prime {\bf 1}/c_i(\lambda) = E(  \gamma( Y(\xvec{s}_1)^2 ) )/c_i(\lambda)$ and thus
		\[ \kappa_i = \frac{\kappa c_i(\lambda)}{c_i(\lambda) +\kappa E(  \gamma( Y(\xvec{s}_1)^2 ) } = \frac{\kappa}{1+\kappa (1-\lambda)  E(  \gamma( Y(\xvec{s}_1)^2 ) )} \]
		does not depend on $i$ at all. We briefly write $\kappa_1$.
		
		Then the left side of (\ref{limit5}) is equal to
		\[ \frac{1}{n} \sum_{i=1}^n \left( \mbox{log}(1 + \kappa Z_{in} ) - E(\log(1 + \kappa Z_{in}) \right) \]
		\[ = \frac{1}{n} \sum_{i=1}^n \left( \log\left(1 + \kappa_1 (Z_{in} - E(Z_{in})) \right) - E\left( \log\left(1 + \kappa_1 (Z_{in} - E(Z_{in})) \right) \right) \right) \]

		\[= I_n - E(I_n) . \]
		Using (\ref{log}) we get that $I_n = II_n  + III_n$ with $II_n = \frac{\kappa_1}{n} \sum_{i=1}^n ( Z_{in} - E(Z_{in}) )$ and
		\[ | III_n | \le  IV_n = \frac{\kappa_1^2}{n} \sum_{i=1}^n  ( Z_{in} - E(Z_{in}))^2 ) . \]
		Since
		\[  \sum_{i=1}^n  ( Z_{in} - E(Z_{in}))^2 ) = (1-\lambda)^2 \sum_{i=1}^n ( \xvec{d}_i(\lambda)^\prime \xvec{\Delta}_i )^2 \]
		it follows with  (\ref{limit7}) that $IV_n \stackrel{p}{\rightarrow} 0$. Because
		\[ | \frac{1}{n} \sum_{i=1}^n ( Z_{in} - E(Z_{in}) ) | \le \left(  \frac{1}{n} \sum_{i=1}^n  ( Z_{in} - E(Z_{in}))^2 ) \right)^{1/2} \]
		it holds that $II_n  \stackrel{p}{\rightarrow} 0$ as well and part a) is proved.

		To prove part b) the Markov inequality is applied. It holds that
		\[  P (  \frac{1}{n} \sum_{i=1}^n  ( Z_{in} - E(Z_{in}))^2  > \varepsilon) \le \frac{1}{n} \sum_{i=1}^n  Var( Z_{in} )    \]
		and
		\[ \sum_{i=1}^n  Var( Z_{in} ) = (1-\lambda)^2 \xvec{1}^\prime (\xmat{I} - \lambda \xmat{W}_2^*)^{-1} \xmat{W}_1^* Cov( \xvec{\Delta} ) \xmat{W}_1^{* \prime} (\xmat{I} - \lambda \xmat{W}_2^{* \prime})^{-1} \xvec{1} \]
		and thus the result follows.

	\end{proof}


\end{document}